\documentclass[onecolumn,12pt,draftcls]{IEEEtran}

\usepackage[T1]{fontenc}
\usepackage{graphicx}
\usepackage{cite}
\usepackage{subcaption}

\usepackage{overpic}
\usepackage{amssymb}
\usepackage{amsmath}
\usepackage{array}
\usepackage{color}

\IEEEoverridecommandlockouts

\begin{document}

\title{Reproducible Research: \\ Best Practices and Potential Misuse}

\author{ 
	\IEEEauthorblockN{Emil Bj\"ornson} \vspace{-1cm}
}

\maketitle


\IEEEpeerreviewmaketitle

The scientific world is becoming more open to the public and fellow researchers. Open access publishing is becoming accepted, even if some publishers are resisting. The next step is the open code and data paradigm, which was briefly discussed in the ``From the Editor'' column in the November 2018 issue of IEEE Signal Processing Magazine (SPM) \cite{8498085}. In this column, I follow up on this topic by sharing my experiences, best practices, and thoughts about reproducible research.

During my doctoral studies in signal processing for communications, I realized that some algorithms have been developed multiple times for similar but nonidentical purposes. One key example can be found in the topic of transmit beamforming, which I wrote about in the ``Lecture Notes'' column in the July 2014 issue of SPM \cite{Bjornson2014d}. Suppose we are using an antenna array to beamform a wireless signal toward an intended receiver while simultaneously limiting the resulting interference to a set of nonintended receivers. Because these are two conflicting goals, no beamforming solution can be optimal in both respects. However, one good heuristic algorithm is to first select the signal's beamforming, which is represented by a vector, as a matched filter to the channel of the intended receiver. Next, the vector is multiplied with the regularized inverse of the Gram matrix of the nonintended receivers' channel vectors. This matrix inverse is a type of whitening filter, but its details are not important in this context. I discovered that many different authors, using slightly different motivations, have independently proposed this heuristic algorithm over the past two decades without referencing one another. The heuristic has been given many names, including the transmit Wiener filter, signal-to-leakage-and-noise ratio beamforming, transmit minimum mean-square error beamforming, and virtual signal-to-interference-plus-noise beamforming. How could this happen?

It is easy to blame situations such as this on poor literature investigation and conclude that researchers should spend more time reading and reflecting on papers and less time writing new ones. Although I sympathize with that sentiment, I do not think it can reasonably resolve the situation. The earliest paper on the beamforming heuristic that I have found is from 1995. At that time, a literature search meant trekking through the dusty library stacks \cite{8498085}. When I began my doctoral studies in 2007, I could use IEEE Xplore to search for papers, which was much more convenient. However, the number of papers on transmit beamforming had grown tremendously in the meantime. The large amount of interest in signal processing for 5G communications has made the situation even more unmanageable for students who are entering the field today. Although online access to papers was certainly a leap forward, it did not solve the volume problem. For instance, IEEE Xplore contains thousands of papers on transmit beamforming, so reading all of them just to check whether the algorithm that you just derived already exists is not reasonable. Currently, it is possible to search the literature databases for only keywords and not assumptions, logical arguments, or mathematical expressions. Perhaps machine learning will eventually come to our aid, but it has not yet.

I believe the core issue is that researchers, including me, generally formulate a research problem, browse the last few years of published papers in a small set of venues to verify that authors in these conferences and journal issues have not solved this specific problem, and, finally, develop a new tool (e.g., an algorithm) that solves the problem. The more unique the solution is, the easier it is to publish. However, after decades of signal processing research, the community has already developed a variety of powerful algorithms to apply to many different problems. Hence, when I formulate a new research problem, there is a good chance that a similar one was already considered decades ago. For instance, how many beamforming researchers of today are familiar with the survey paper from 1988 \cite{Veen1998a} that summarizes key results from the 1970s and 1980s? It is even plausible that a problem that is conceptually different but mathematically similar to the one you are currently studying has been solved in a very different field; for example, there are striking mathematical similarities between beamforming optimization in communications and portfolio optimization in financial engineering \cite{SIG-072}. If we better apply the collective knowledge and insights from the past, our community can make swifter progress and tackle problems that are more challenging.

\section*{The Reproducibility Crisis}

Signal processing research is a hybrid of theoretical and experimental/computational research \cite{Kovacevic2007a}, \cite{Vandewalle2009}. The community is excellent with theory, for example, in terms of developing rigorous theories for estimating, detecting, and optimizing signals and systems. This part is published in our papers as lemmas and theorems, followed by detailed proofs that are (hopefully) carefully scrutinized in the peer-review process. Readers of published signal processing papers can continue to examine these parts after publication and make corrections when needed.

The experimental/computational part is typically represented by a section with numerical simulations, where a new signal processing algorithm is shown to outperform prior work. Unfortunately, this part is far behind the theory part in terms of scientific rigor. Anecdotal evidence is the rule rather than the exception, at least in signal processing for communications: the authors usually select one system setup (perhaps randomly or possibly by cherry picking) and apply their new algorithm to it. If the authors can show substantial gains as compared with a previously published algorithm, they can claim the superiority of the new algorithm, and the reviewers generally do not complain. However, this is a poor practice for several reasons. First, the unpaid reviewers generally do not have the time or resources to validate the accuracy of the numerical simulations or experimental results; as long as the results are reasonable, the reviewer has no choice but to trust that they are correct. Second, it is fully possible, even likely, that researchers can find another setup in which the previous algorithm outperforms the new one. In fact, the authors are probably comparing their new well-tweaked algorithm with their quick implementation of the competing algorithm, which may be full of bugs.

In experimental and computational research, the reproducibility of published results and the use of large common data sets for evaluation are essential for building confidence and drawing accurate conclusions. For example, in the topics of biology and life science, it is not until a result is independently reproduced by other researchers that it is considered valid \cite{Vandewalle2009}. In our field, the measurement data and algorithmic implementation contribute as much as the analysis, but this information does not fit within the restrictive format of papers. Unfortunately, the simulation code and data are traditionally kept secret, which creates a competitive advantage for the authors of a paper to continue ``their'' line of research because other researchers need to spend excessive time reimplementing state-of-the-art methods before attempting to advance them. Even if they do their best to reproduce published results, they might fail. In a large Springer Nature study, more than 70\% of the researchers surveyed were unable to reproduce other researchers' results \cite{Baker2016a}, and 50\% of them believed that there is a significant reproducibility crisis. In fact, many researchers even fail to reproduce their own prior results \cite{Vandewalle2009}. The survey considers many different research fields, including engineering. When a figure in a paper cannot be reproduced, it does not necessarily mean that the entire paper is wrong or that the researchers have consciously manipulated anything. Forgetting to write out one parameter value in a computer simulation \cite{Vandewalle2009} or accidentally including the incorrect version of a graph is enough to render the results of a paper nonreproducible. I have personally made these mistakes!

The journals in our field expect novel contributions while papers that validate previous results might face immediate rejection. The IEEE Signal Processing Society's (SPS's) ``Information for Authors'' page states that papers containing ``a straightforward combination of theories and algorithms that are well established and are repeated on a known scenario'' should be rejected. This makes sense for the theory part of our research (although new, shorter proofs of known theorems also have an important value) but not for the experimental part because its scientific value builds on the trust achieved through reproducibility. As long as it is not mandatory to publish simulation code with a paper, we should encourage people to reproduce each others' results and publish their findings.

\section*{Best Practices}

\begin{figure}[!t]
\centering
\includegraphics[width=\columnwidth]{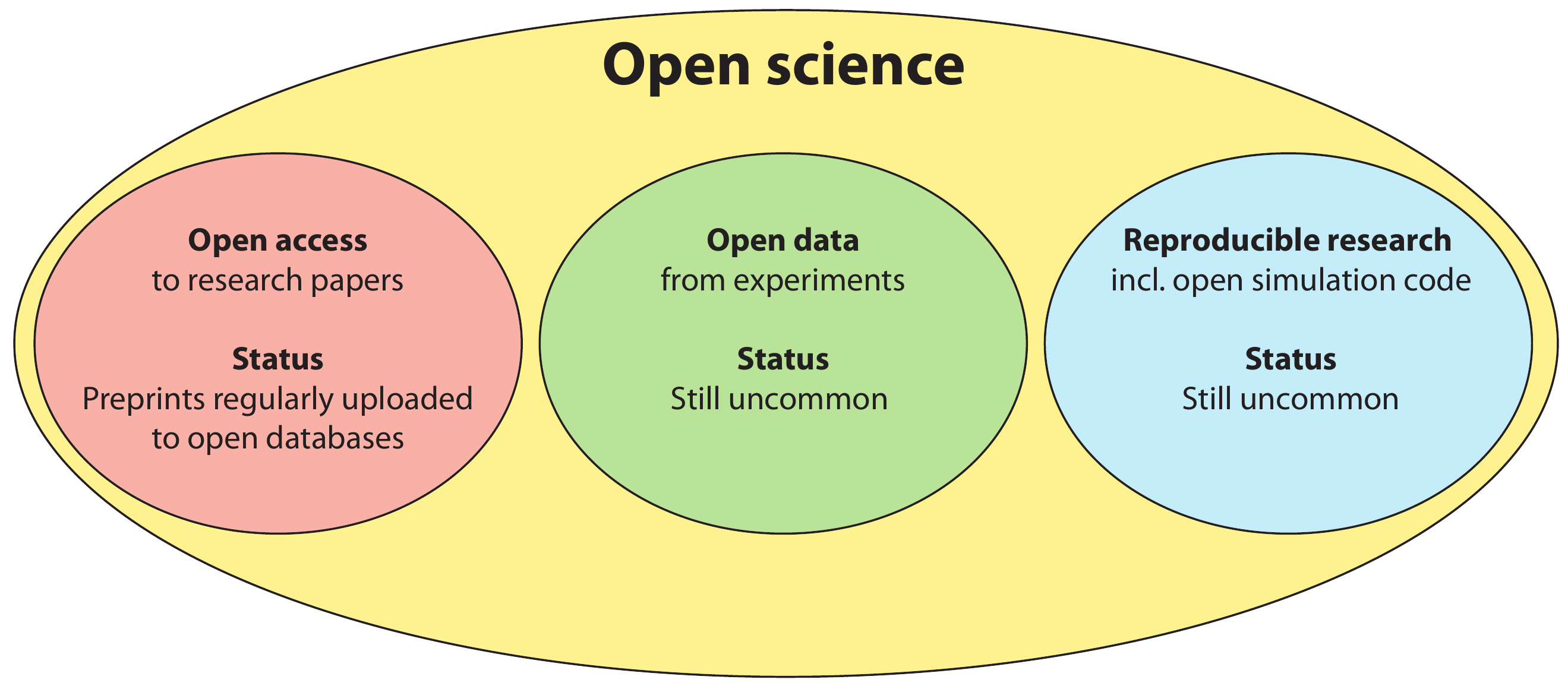}
\caption{The taxonomy of open science specifies three key components: open access, open data, and reproducible research.}
\label{open-science}
\end{figure} 

The issues with reproducibility have been noted in the SPS before. Barni and Perez-Gonzalez wrote an excellent SPM column with the controversial title ``Pushing Science Into Signal Processing'' in 2005 \cite{Barni2005a}. It inspired the special session ``Reproducible Signal Processing Research'' at the ICASSP 2007. A paper from that session, ``How to Encourage and Publish Reproducible Research'' by Kovacevic \cite{Kovacevic2007a}, first made me aware of the reproducibility crisis and necessary movement toward open code and data. However, Claerbout and Karrenbach \cite{Claerbout1992} are early proponents of reproducibility. In 2009, Vandewalle et al. \cite{Vandewalle2009} followed up on the ICASSP session with their article ``Reproducible Research in Signal Processing,'' and stated that research is reproducible when ``all information relevant to the work, including, but not limited to, text, data and code, is made available, such that an independent researcher can reproduce the results.''

Research reproducibility is one of the components of open science. In Figure~\ref{open-science}, I provide a taxonomy, where the three key components are open access to research papers, open data from experiments, and reproducible research (including open simulation code). In recent years, we have progressed quite far on the first one but not the other two.

Inspired by \cite{Kovacevic2007a}, \cite{Vandewalle2009}, and \cite{Barni2005a} and their suggested procedures, I have made the simulation code and data for 30 publications openly available online. It began with my first textbook \cite{Bjornson2013d}, which I wrote to explain how the same set of optimization algorithms for communications have been applied to a variety of seemingly different applications in signal processing. Because I was sure that new applications would arise where these algorithms could be readily applied, it was natural to provide reference implementations to encourage reuse over reinvention. My largest project so far is the 500-page textbook Massive MIMO Networks: Spectral, Energy, and Hardware Efficiency \cite{massivemimobook}, in which every single simulation figure can be reproduced by using the accompanying MATLAB code. All of the code has been published under the GNU General Public License, which gives users the right to run and modify the code as well as share modified versions using the same license. I chose this license mainly due to laziness---it is widely used---but \cite{Vandewalle2009} describes alternatives with special features for research purposes.

If you make your code available online, there is a good chance that someone will download it and try to understand every single line. This is what scared me in the beginning: What if someone finds a critical error in my code? This fear eventually made me realize that if my code is not good enough to be scrutinized, why would my papers deserve to be published? I therefore decided to sharpen the experimental part of my research. At the point when I would have previously submitted a paper, I now instead take the following steps:

\begin{enumerate}
\item Copy the code into an empty folder.
\item Remove all unnecessary files.
\item Write detailed comments in every file and check that parameter values and algorithms are exactly the same as described in the paper.
\item Delete unused features.
\item Set a ``random'' seed to the random number generators.
\item Write instructions for how to run the code.
\item Run the code according to the instructions to verify that it actually reproduces the plots in the paper.
\end{enumerate}

Many mistakes can be discovered and corrected by following this procedure. Even when the code is technically correct from the beginning, there can be reproducibility issues due to discrepancies between what is done in the code and what is written in the paper. For instance, parameter values may be different or missing, or there may be undocumented features (e.g., for the purpose of numerical stability or algorithmic initiation/termination). Also, the code may use an insufficient number of Monte Carlo realizations to obtain accurate results, or external data may be used without stating how it was preprocessed (e.g., which size and color format the analyzed image had \cite{Vandewalle2009}). These things are easy to miss unless you are analyzing the code systematically. The first time you follow the suggested procedure to validate your code, it will certainly take additional time, but you will save time, in the long run, because you (and others) can more easily reuse and revise the code in future publications. I strongly recommend tidying up the code right before submission instead of at the time of publication, both to minimize the chance that errors reach the submitted version and because this is the time when you will best remember the details.

A change toward open code and data will not happen automatically, but the SPS needs and deserves a strong push in the right direction. This can be achieved by either promoting good practices or having strict requirements. For example, Springer Nature has detailed research data policies in which some of its journals encourage the publication of data and others require the underlying data to be published and peer reviewed. Many research-funding agencies require open access publishing and may soon add open access for data and code to their requirements. In 2007, Kovacevic \cite{Kovacevic2007a} suggested a course of action for the SPS. In addition to publishing code along with papers, she suggested encouraging the publication of more experimental work and the analysis of known algorithms in new settings or with different data. She also suggested promoting the value of such work by dedicated special issues, paper awards, and student training. Unfortunately, not many of these suggestions have been implemented so far, but at least the IEEE is providing functionality for sharing data (such as IEEE DataPort) and simulation code (Code Ocean) in its ecosystem. Personally, I am sharing my data and code on GitHub because I started sharing there before the IEEE released its alternatives.

From my experience, there are four important things to consider when sharing code or data. First, it should be easy for the reader to become aware of the simulation code's existence, which means that authors must state it explicitly in the paper. I started by adding a footnote on the first page and in IEEE Xplore, but then I switched to stating it in the simulation section. I now include a statement in the introductory section to give it the same visibility as the list of mathematical notations. Second, the code/data should be openly and permanently available, which means not putting anything on your current website or requiring people to send an email to your current address. Use a free and reputable repository, preferably hosted by a not-for-profit organization: something like arXiv.org but for code and data.

Third, make sure that you can revise the code over time in case bugs are discovered. During the first year that my book \cite{massivemimobook} was available, several readers contacted me with detailed questions on the code, which allowed me to correct a few bugs and upload a revised version. Fourth, declare which software is needed to run the code, including the version and operating system you used. The list should preferably contain only nonproprietary software; for example, if you are coding in MATLAB (as I am), you can try to make the code compatible with GNU Octave. However, there are cases in which it is necessary to use commercial software toolboxes to produce faster research progress, just as there are cases when you need particular hardware capabilities to run the code (e.g., some of the simulations in my book \cite{massivemimobook} require so much memory and processing power that we ran them on a supercomputer). In any case, it is likely that the required software will eventually become outdated, but we should do our best to enable future reproducibility.

When it becomes standard practice to publish code and data, there will certainly be a risk that we will quickly become flooded by code; if there are hundreds or thousands of papers with accompanying code on a topic, it is not reasonably possible to go through all of them to decide what to reuse or what to compare your current research against. Even if you find the code for the most relevant prior works, it may be implemented in different programming languages or with special properties, which makes interfacing as difficult as actually reimplementing the methods from the beginning. Therefore, a common library with reference implementations will eventually be necessary. It can initially be delivered as supplementary material to textbooks (as I tried with \cite{massivemimobook}), but it should preferably become a library in some mainstream programming language such as Python. We can then finally separate the development of general-purpose algorithms in signal processing from applying them to solve new problems.

What if your paper builds on confidential data or commercial implementations? In the former case, the data can be filtered to remove the confidential parts, and the analysis should (as far as possible) be reproducible by using the filtered data. In the latter case, a company may have spent years developing the simulation environment they use in their paper and want to keep it proprietary. In other words, the paper is written to advertise the company's expertise rather than make a real scientific contribution. One viable solution is requiring that the simulation code is shared with all journal publications, but not when publishing at conferences. Therefore, if you do not want to share your code and data, then only publish conference papers. This is consistent with my work practice: I treat conference papers and journal preprints as preliminary work, and then I make the code publicly available when the journal paper is accepted for publication.

\section*{Potential misuse}

The openness and reproducibility that I advocate can also be misused. Rather than inspiring more substantial and provable scientific progress, the ``publish or perish'' pressure to publish many papers can also lead to more ``e-improvement'' papers. Existing code can be slightly modified and applied to a slightly different setup, leading to a new paper. There is a thin line between such a practice and actually publishing more experimental verifications of known algorithms, as I suggested before. Maybe there is no way to separate one from the other?

The more troubling issue is plagiarism, which becomes much easier to perform when there is open access to papers, code, and data. Did you know that anyone can download the source code of papers that are uploaded to arXiv.org? One can easily take the existing text and equations from a paper, modify the sentence structures, and replace some words with synonyms to fool plagiarism-detection software. This is known as rogeting, as a reference to Roget's Thesaurus. Next, the person would download the simulation code, change a few parameter values, and generate new plots. With under an hour of work, you can produce a paper that looks novel, and if you choose to plagiarize an obscure paper, even an expert reviewer might not recognize the plagiarism. In 2018 alone, I was asked to review three papers that plagiarized my work; I am sure other cases slipped through their review processes.

There are no easy ways to prevent plagiarism. As we refine the procedures for plagiarism detection, new ways to fool the system will arise. Perhaps we can learn something from the sports world: antidoping laboratories are saving blood samples from the Olympics and can retest them 10 years later, when the ability to detect certain doping substances has improved. Similarly, when the automatic plagiarism-detection software is refined, we can use it to reanalyze previously published papers to look for plagiarism that slipped through. When someone is caught for plagiarism, he or she should be suspended not only by one publisher (as was the case in the three occurrences that I experienced) but be blacklisted by all international publishers that want to be taken seriously, as is done in the sports world.

\section*{Conclusions}

We have taken steps toward a more open scientific environment in recent years, but much work remains. The current review process is designed to validate the theoretical part of signal processing papers, but the experimental part deserves the same attention. To enable this, every author should share the code and data underpinning their journal papers, first with the reviewers and, later, with fellow researchers, to continue to review and improve the results. A standard procedure to achieve reproducibility is defined in \cite{Vandewalle2009}, and I have outlined my workflow previously in this article. Research reproducibility really is a win--win situation: authors can spend more time being innovative and also benefit from more citations \cite{Piwowar2007}. Openness and transparency are also important for emphasis in a time when personal opinions receive the same or higher priority as expert knowledge in the media.

In addition to enhancing the research reproducibility of individual papers, we can improve scientific progress by building common code libraries for signal processing algorithms. The classic algorithms for filtering and the spectral analysis of signals exist in many programming languages but can be complemented with more advanced methods that are not taught in undergraduate classes yet are standard tools in research. In many cases, a paper's algorithmic contribution is tweaking an existing algorithm to fit a new problem. Therefore, there is no need to reimplement it from scratch, but one can preferably inherit and improve upon a standard implementation. By storing both the basic algorithms and various extensions in the same common environment, scientific progress is easier to track, published methods are easier to compare, and relevant prior works are easier to identify---which was the initial problem mentioned in this article.

\section*{Author}

\textbf{Emil Bj\"ornson} (emil.bjornson@liu.se) is an associate professor at Link\"oping University, Sweden. His research interests are signal processing, multi-antenna communications, energy efficiency, and machine learning. He is the first author of the textbooks \cite{Bjornson2013d} and \cite{massivemimobook}. He is an elected member of the Signal Processing for Communications and Networking Technical Committee (SPCOM TC) and currently serves as an associate editor for IEEE Transactions on Communications and IEEE Transactions on Green Communications and Networking. He received the 2018 Marconi Prize Paper Award in Wireless Communications and best paper awards at WCSP 2017, IEEE ICC 2015, IEEE WCNC 2014, IEEE SAM 2014, IEEE CAMSAP 2011, and WCSP 2009. He is dedicated to reproducible research and has made a large amount of simulation code publicly available.

\bibliographystyle{IEEEtran}
\bibliography{IEEEabrv,refs}

\end{document}